\documentclass[sigconf]{acmart}

\usepackage{booktabs} 
\usepackage{xspace}

\usepackage[ruled]{algorithm2e} 

\SetAlFnt{\small}
\SetAlCapFnt{\small}
\SetAlCapNameFnt{\small}
\SetAlCapHSkip{0pt}
\IncMargin{-\parindent}

\acmJournal{PACMHCI}
\acmVolume{9}
\acmNumber{4}
\acmArticle{39}
\acmYear{2010}
\acmMonth{3}
\acmArticleSeq{11}

\newcommand{\eg}{e.\,g.\xspace}

\begin{document}
\title{ARCHANGEL: Trusted Archives of Digital Public Documents}

\author{J. Collomosse, T. Bui, A. Brown}
\affiliation{%
  \institution{University of Surrey}
  \city{Guildford, Surrey, UK}
}
\email{j.collomosse@surrey.ac.uk}
\author{J. Sheridan, A. Green, M. Bell}
\affiliation{%
  \institution{The National Archives}
  \city{Kew, London, UK}
}
\email{jsherdian@nationalarchives.gov.uk}
\author{J. Fawcett, J. Higgins \\O. Thereaux  }
\affiliation{%
  \institution{Open Data Institute (ODI), UK}
}
\email{jamie.fawcett@theodi.org}

\begin{abstract}

We present ARCHANGEL; a de-centralised platform for ensuring the long-term integrity of digital documents stored within public archives.  Document integrity is fundamental to public trust in archives.  Yet currently that trust is built upon institutional reputation --- trust at face value in a centralised authority, like a national government archive or University.  ARCHANGEL proposes a shift to a technological underscoring of that trust, using distributed ledger technology (DLT) to cryptographically guarantee the provenance, immutability and so the integrity of archived documents.  We describe the ARCHANGEL architecture, and report on a prototype of that architecture build over the Ethereum infrastructure.  We report early evaluation and feedback of ARCHANGEL from stakeholders in the research data archives space.
\end{abstract}

%
%


%
%



\maketitle

\renewcommand{\shortauthors}{J. Collomosse et al.}

\section{Introduction}

Archives and Memory Institutions (AMIs) are the lens through which future generations will perceive today. AMIs are founded upon the principles of public trust --- of being neutral and completely trustworthy. The immutability and integrity of the documents they hold are essential to maintaining their objectivity; be they government documents in  National Archives, or research documents held by University archives. Yet, today's digital age presents urgent, new challenges to this trust and immutability.  Digital documents are ephemeral, and produced in great volume.  Their intangibility leaves them open to modification --- not only to tampering but also to degradation over longitudinal time periods (\eg decades).  For example, file formats become obsolete and documents are transcoded. How can we ensure that meaning is not lost?

In this paper we describe ARCHANGEL; a de-centralised architecture for ensuring the integrity of digital documents within public archives.  ARCHANGEL utilises blockchains --- a form of secure decentralised ledger technology (DLT) --- as a basis for ensuring the provenance and integrity of documents.  Blockchains store chronologically ordered transactional data, permanently preserving that data through peer-to-peer distribution and consensus checking without the need for a trusted third party.  Although best known for underpinning digital exchanges of cryptocurrencies \eg BitCoin \cite{Narayanan2016}, their transaction based model has also been applied to transmit data payloads for internet domain name management (Namecoin) \cite{Kalodner2005} and interpersonal messaging (Bitmessage) \cite{Buterin2012}. Recently there has been significant interest in applications of DLT to public services by government \cite{Walport2015} including to record-keeping \cite{Lemieux2016}. ARCHANGEL breaks new ground by proposing the use of a blockchain payload to record digital signatures ({\em content evidence}) derived from either scanned physical, or born-digital, document to ensure their integrity over decade- or century-long timespans.

We have implemented a prototype of ARCHANGEL based on the Ethereum DLT infrastructure and exposed this to archivist stakeholders in the University research data management space for feedback. We document this prototype in Sec.~\ref{sec:main}, and reflect upon this early evaluation in Sec.~\ref{sec:eval}.  We discuss future directions for development of the ARCHANGEL platform in Sec.~\ref{sec:conclude}.

\section{Related Work}
\label{sec:bg}

AMIs are struggling to keep pace with the exponential rate of digital transformation in  society \cite{TNA2014,Walport2015}.  Today, the majority of content is born-digital and there is inexorable end-user demand for digitisation of existing content e.g. for open access and operational efficiencies \cite{Johnson2013}.  Recent work addressing this trend focuses on developing theories of record keeping, such as the Records Continuum Model \cite{McKemmish2010}, or standards and technologies for describing, cataloguing or searching archives such as Discovery \cite{Cates2014}, the Archives Portal Europe project \cite{APE} or the Records in Context initiative \cite{Cates2016}. Some work has been done looking at analytics of archival data enabled by big data approaches such as Traces Through Time \cite{Bell2015}. Rather than exploring novel ways to index and annotate documents with metadata, ARCHANGEL explores the orthogonal challenge of ensuring the long-term integrity and sustainability of digital archival content, proposing a platform for verifying the integrity and provenance of digital documents whilst entrusted to the archive (curation) and upon document release (presentation).

AMIs now exist within an age where people are increasingly questioning institutions and their legitimacy. Historically, an archives' word was authoritative, in effect vouching for the integrity of documents through their reputation.  ARCHANGEL advocates the use of a de-centralised model (DLT) rather than --- say --- a secondary database of  document signatures or certificate key authorities maintained centrally, to evidence the integrity and provenance of those documents \cite{Cates2016}.   In doing so, ARCHANGEL enables a shift from an institutional underscoring of trust, to a technological underscoring of trust.  Some recent work has explored distributed assurance of provenance through embedding document signatures within versioning information \eg in Microsoft Word documents \cite{Shatnawi2017,Filho2017}.  ARCHANGEL goes further, creating a de-centralised repository of content evidence independent of the document itself that is collectively maintained across multiple participating archives through consensus checking on the Blockchain.

\section{Trusted Digital Documents}
\label{sec:main}

ARCHANGEL utilises a permissioned blockchain model, in which operators or automatic processes authorised to add content to the AMI commit blocks into the chain encoding content evidence (and if that evidence is derived in a non-standard way, a hash of associated code for deriving that evidence).  The security of a blockchain is afforded by the immutability of data with the blocks, delivered by the compounding effect of each new block being hashed to include the hashes of previous blocks.  Thus as content is committed into the blockchain, the security of the content is reinforced.  The Blockchain remains publicly readable to enable open verification of documents released from the archive at any time.

ARCHANGEL proposes a cross-AMI model in which a single DLT is contributed to by multiple AMIs, potentially across different disciplines and nations, mitigating any risk of distortion of an archive by its operating AMI.  In this way multiple archives potentially across international borders, can mutually reinforce the integrity of each others' documents without necessarily being party to the content of those documents until release.

\subsection{Illustrative Scenarios}
\label{sec:scenario}

Documentary evidence gathered for public inquiries (\eg into the UK 7/7 terrorist attacks, or the Chilcot Inquiry) are kept by the UK National Archives and can remain closed for the best-part of a century. ARCHANGEL enables such archives to lodge content evidence within the Blockchain at time of document deposition, enabling the public to verify the integrity and provenance of those documents upon release.  The verification is open and may be performed by anyone through reapplying the hashing algorithms used to extract content evidence at deposition, and comparing those hashes to those lodged within the Blockchain also at the time of deposition.  If a non-standard hashing algorithm (for example a machine learning based content parser) was used, then the hash of that algorithm code may also be included in the Blockchain.

Similarly, consider a University publishing a study on climate change and including a DOI to supporting data released openly.  That data could be hashed into content evidence, and lodged in the ARCHANGEL DLT alongside a hash of the code necessary to recreate that hash at time of publication.  A decade passes, and years later the research integrity of the paper is called into question.  The University can evidence that the research data it provides matches that at time of deposition.

\begin{figure}[t!]
\centering
\includegraphics[width=0.9\linewidth]{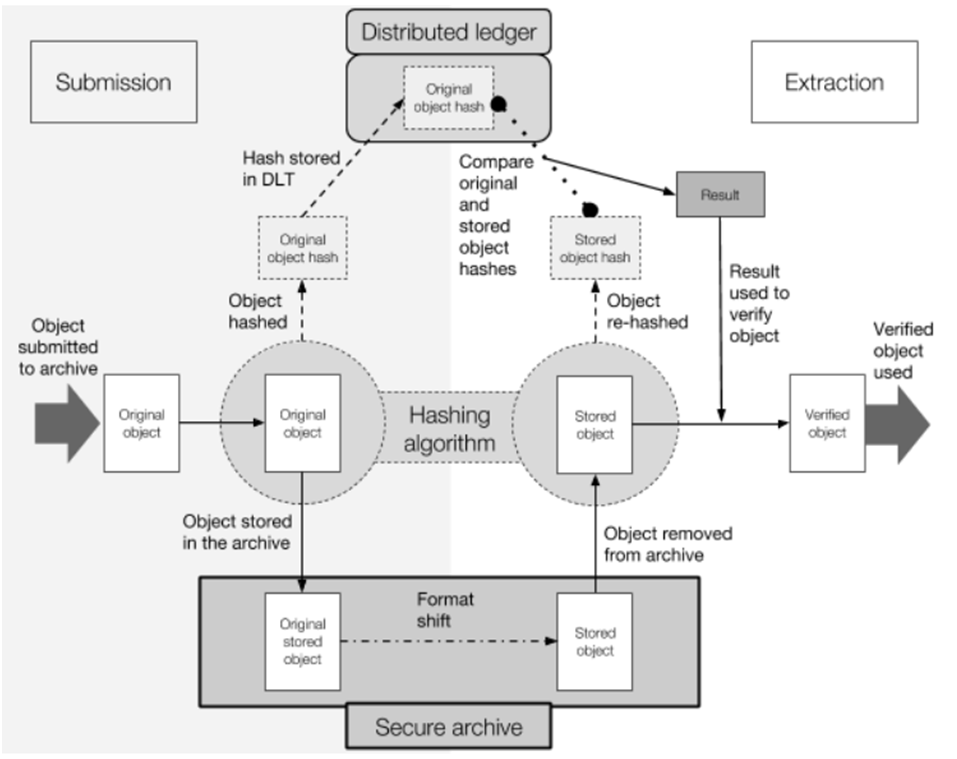}
\caption{Architecture of the proposed ARCHANGEL platform.  Documents are processed to extract {\em content evidence} which is stored immutably within a Blockchain alongside metadata identifying both that content, and the algorithm used to extract the evidence. A document's integrity and provenance can be checked at any time by re-extracting and comparing the content evidence to that in the Blockchain.   \label{fig:arch}
}
\vspace{-10pt}
\end{figure}

\subsection{ARCHANGEL Architecture}
\label{sec:arch}

Figure~\ref{fig:arch} illustrates the ARCHANGEL architecture.  ARCHANGEL does not propose a distributed filesystem or similar for the storage of documents (the AMI is assumed to provide this solution) rather we propose the decentralised storage of compact hashes derived from documents on a Blockchain alongside metadata to assist in future identification and verification of those documents.

Upon deposition of a document, a file format identification tool determines the content type of the document (\eg PDF, Microsoft Word) by performing classification upon the binary information within the file irrespective of its accompanying metadata \eg filename.  Content evidence is then extracted from the document in a format-dependent manner via a content hashing algorithm.  In the simplest form this content hashing might be a classical binary hashing algorithm (\eg SHA-256) applicable to all formats, however we consider that bespoke content hashing processes might be applicable to specific formats. For example, a digital image of a scanned physical document might employ a deep neural network (DNN) to extract robust visual features from visual content that are invariant to appearance properties (\eg illumination, ageing) of that document. 
Having extracted the content evidence, a filename or other global unique identifier (GUID), the content hash, and an unique identifier signifying the content hashing process are stored alongside supplemental metadata with the Blockchain.  This supplemental metadata might include archivist's notes, date of deposition, versioning information, and if the content hashing process is bespoke potentially a hash (using a binary standard hash such as SHA-256) of the code or equivalently DNN model used to extract the features.  In the latter case, the code or model would also be secured within the archive.

To store this block of new data, it is appended to the end of a linked list of blocks in a distributed data structure (the 'Blockchain').  Multiple nodes within the DLT infrastructure replicate this operation and consensus check the result per the protocol of the underlying DLT infrastructure.  We have chosen to implement ARCHANGEL in Ethereum DLT infrastructure (c.f. Sec.~\ref{sec:proto}), which is currently a proof-of-work based Blockchain.  In practice this means that multiple nodes in the network must solve a cryptographic puzzle in order to append to the list while other nodes must be in consensus as to the validity of the operation \cite{Narayanan2016}. In our architecture we propose two modes of consensus checking, both predicated upon a permissioned DLT model:

\begin{enumerate}
    \item {The Blockchain is maintained via proof of work across a private set of nodes, which are maintained collectively by multiple AMIs each with independent governance structure \eg national archives of different nation states.  As such an unprecedented level of collusion would be required to corrupt the Blockchain.}
    \item {The Blockchain is maintained via proof of work across a public Blockchain maintained globally.  In such case a program embedded within the Blockchain (a 'smart contract') with sole permission to write to the Blockchain is invoked in order to the append data.  Access to the smart contract end-point is granted via secret key.  In this case corruption would require more than half of the public DLT infrastructure miners to collude, which is again unlikely \eg on the Ethereum main network. }
\end{enumerate}

To verify the provenance and integrity of a document curated or released by an AMI, the content evidence must be computed from that copy of the document and compared against that immutably stored within the Blockchain.  The public availability of the contents of the Blockchain enable anyone to search and identify the appropriate data block using the unique identifying information (GUID) accompanying the content hash, stored during deposition. Should the content evidence be hashed using a bespoke technique, the instance of that technique (code or network model) must too be requested from the AMI and compared to the hash computed at deposition.  Assuming both content and algorithm hashes match, the document is considered to be authentic.

\subsection{Prototype Implementation}
\label{sec:proto}
\begin{figure}[t!]
\centering
\includegraphics[width=1.0\linewidth]{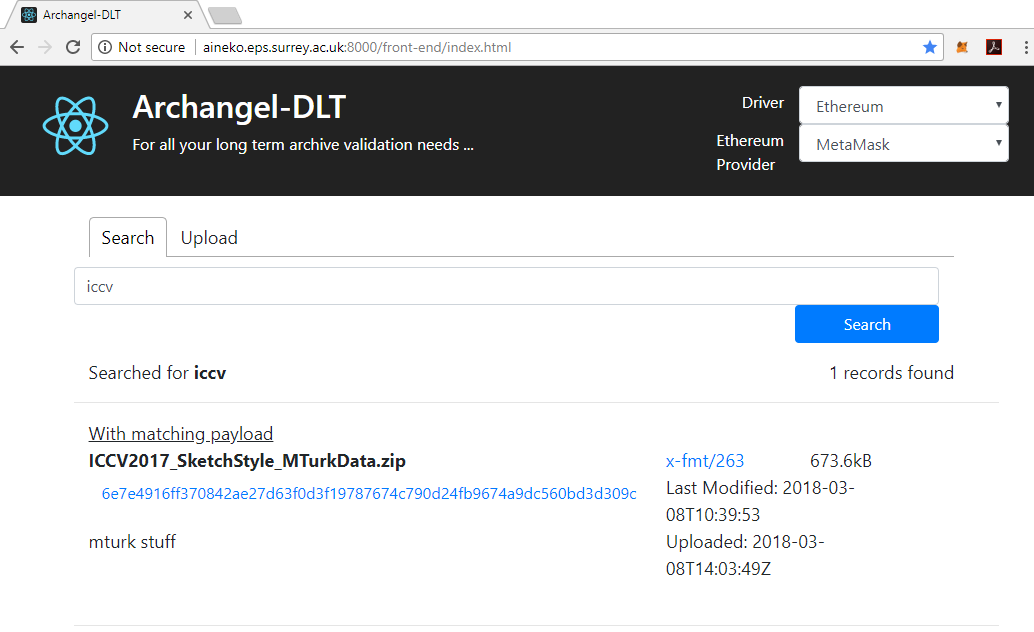}
\caption{Screenshot of the ARCHANGEL platform running on Ethereum  (search functionality shown).  User may search on document global unique identifier (GUID), metadata, or content evidence (hash) to verify its integrity.   \label{fig:screenshot}
\vspace{-10pt}
}
\end{figure}

We have implemented a prototype of ARCHANGEL on the Ethereum public test net (Rinkeby), adopting consensus model (2) described above.  Ethereum was selected due to its global adoption and prominence as a DLT platform, and due to its technical capability to store both data and execute smart contracts (via EVM); which we implemented in the Solidity language.  

We used the DROID (Digital Record Object IDentification) application developed by the UK National Archives to classify document type \cite{DROID}.  Without loss of generality we assume that this process is executed via a cloud-based service capable of accepting uploads and running them both through DROID and through content extraction.  Currently our prototype uses a standardised binary hash (SHA-256) but ongoing work is exploring format-specific hashing using bespoke machine learning models for document feature extraction.

Figure~\ref{fig:screenshot} presents a screen-shot of our implementation which contains the functionality to 1) to deposit documents; 2) to search for documents \eg based on GUID or content hash string; and 3) to verify documents (effectively running operations 1) and 2) in succession to identify prior instances of the content hash).

\section{Stakeholder Workshop}
\label{sec:eval}

A workshop was convened with 13 expert participants drawn from AMIs in the government, commercial legal and university research document management spaces.  Participants were briefed on the ARCHANGEL platform and provided the opportunity to interact with the platform prototype for one hour in a lab-based setting given a set of research documents provided by the University of Surrey.  Participants explored the functionality of the platform to deposit, search and verify documents not as a usability exercise but as a provocation for a facilitated semi-structured group discussion to capture feedback on the value of ARCHANGEL to digital document preservation in AMIs.  The dominant theme was the perceived value in enabling archives to engender trust both in their records and in their practices; that as a result of providing proof that the records have not changed, their authenticity can be demonstrated. Several discussion themes emerged.\\
\noindent {\bf 1. Defenders of the record}.  
In an era when the technologies to fake digital content are becoming increasingly pervasive, it is not surprising that the public has less trust in all things digital. Coupled with a perceived lowering of trust in public institutions and we have a perfect storm for the digital archive. Blockchain offers a shield which archives can use to defend the records as authentic. By enabling researchers to compare the content evidence (including the checksum) of the record to that recorded on the Blockchain, they can see proof that no changes (deliberate or accidental) have been made to the record since it was preserved in the archive.  It was also noted that acceptance of content evidence might eventually become similar to acceptance of DNA evidence in court, but that establishing that level of confidence would require strong public engaged to explain Blockchain in an accessible manner particularly explaining why one could trust the cryptographic assurances inherent in a DLT solution.  It was noted that although this could be considered a form of digital forensics there was no specific precedent or case law (within the UK) in relation to verification of digital documents in AMIs.  There was strong potential for further socio-technical research around the consequences of ARCHANGEL as a technical platform not only in relation to law but also in the potential to evolve archival practice itself.\\
\noindent {\bf 2. Engagement with emerging technologies}. Archives are not generally viewed as digital institutions nor archivists as a digital profession. Yet, there are many areas where archives are actively engaged with digital technology, from the preservation of born-digital records to researching the uses of machine learning for appraisal, selection and access. AMI involvement in the practical application of Blockchain  demonstrates an interest in and openness to new technologies. The attendees were very keen to be kept informed of the project's progress congiscent of the tidal wave of digital documents arriving in their archives at increasing rate.\\
\noindent {\bf 3. Demonstrates our willingness to be transparent in our practices.} While blockchain encourages citizens to trust the records in our custody as being the 'ground truth' it can also make our practices transparent in a way that can be verified by researchers. Archivists would be able (in an automated environment) to use the technology throughout their processing of the records. They could record the content evidence on the blockchain after each significant curatorial action, creating an audit trail of those actions and a series of hashes to verify. This further encourages trust in archives' role as custodians but also in use of best practice. \\
\noindent {\bf 4. Demonstrates the collaborative nature of the digital archives community.} A major benefit for the archival community of using blockchain technology is that it requires collaboration. The ability and willingness of archives to engage with each other as well as other heritage organisations both nationally and internationally is key to its success. Archives have a good track record of supporting each other. The value of a distributed approach to assuring trust in digital records may be most keenly felt by archives at risk. Blockchain provides a way for archives to underscore trust in each others' collections introducing an entirely new form of collective defence of the archive. For example ARCHANGEL raises the potential for AMIs to collaborate to provide the computational power and the consensus checking of the platform --- potentially across international or jurisdictional boundaries. This gives the technology the huge advantage of not only preventing individual governments from tampering with the public record but also guarantees a degree of longevity due to the legislative position of public archives. The technology itself engenders trust by guaranteeing that the records have not been tampered with, and it also allows archives to make their processes transparent both of which encourage trust in their integrity as custodians of the public record.

\section{Conclusions and Next Steps}
\label{sec:conclude}

We have described ARCHANGEL --- a platform for verifying the provenance and integrity of digital documents within public archives.  Uniquely, ARCHANGEL combines content hashing with distributed ledger technology (DLT) to create a de-centralised trust model.  We have proposed an architecture and reported both its implementation on the Ethereum infrastructure, and feedback from an early trial at a focus group of archivists drawn from a research data management background.  Although fully functional, ARCHANGEL is a prototype and several promising directions exist for future extension.  Currently our content hashing is performed using standard binary hashing (\eg SHA-256) and we would like to specialise hashing to particular document types such as PDF or even images and video.  The latter presents the unique possibility of extracting content-aware hashes of scanned documents that are sensitive to tampering or degradation but invariant to factors such as illumination or imaging device.  We are also considering the integration of the W3C proposed PROV standard \cite{PROV} for document versioning given that blocks may only be added to supercede older content (and not deleted) from a Blockchain. Currently our implementation uses smart contracts as a gateway for writing to the Blockchain, but not for search or verification.  We might also explore the use of smart contracts for the latter, exploring new business models to encourage sustainability.  For example, the maintenance of the DLT (in terms of computational effort for mining) might be facilitated by users who seek document verification 'paying' for that service via contribution of mining effort to maintain the DLT.  Our initial feedback from stakeholders is based upon a workshop environment and not a platform deployed in practice.  We believe that ARCHANGEL has the position to disrupt the professional practice of archivists, as a technology tool that enables robust curation of digital documents that can easily suffer damage \eg through format shifting.  We feel ARCHANGEL shows significant promise as a means for ensuring the future sustainability of archives as they undergo the challenges of digital transformation.

\section*{Acknowledgement}

ARCHANGEL is funded by EPSRC Grant Ref: EP/P03151X/1 under the UKRI Digital Economy Programme.

\bibliographystyle{ACM-Reference-Format}
\bibliography{archangel}

\end{document}